\documentclass[pra,twocolumn,eqsecnum,showpacs,amsmath,amssymb,superscriptaddress]{revtex4-1}
\usepackage{graphicx,bm,color,mathptmx,hyperref} 
\newcommand{\ket}[1]{\left\vert #1 \right\rangle}
\newcommand{\bra}[1]{\left\langle #1 \right\vert}

\newcommand{\Tr}{\mathop{\mathrm{Tr}}\nolimits}

\begin{document}

\title{Generation of entangled matter qubits in two opposing
parabolic mirrors}

\author{N. Trautmann}
\affiliation{Institut f\"{u}r Angewandte Physik, 
Technische Universit\"{a}t Darmstadt, D-64289 Darmstadt, Germany}

\author{J. Z. Bern\'{a}d}
\affiliation{Institut f\"{u}r Angewandte Physik, 
Technische Universit\"{a}t Darmstadt, D-64289 Darmstadt, Germany}

\author{M. Sondermann}
\affiliation{Max-Planck-Institut f\"ur die Physik des Lichts,
  G\"{u}nther-Scharowsky-Stra{\ss}e 1, Bau 24, 91058 Erlangen,
  Germany} 
\affiliation{Department f\"{u}r Physik, Universit\"{a}t Erlangen-N\"{u}rnberg,
Staudtstra{\ss}e 7, Bau 2, 91058 Erlangen, Germany}

\author{G. Alber}
\affiliation{Institut f\"{u}r Angewandte Physik, 
Technische Universit\"{a}t Darmstadt, D-64289 Darmstadt, Germany}

\author{L.~L.~S\'{a}nchez-Soto} 
\affiliation{Max-Planck-Institut f\"ur die Physik des Lichts,
  G\"{u}nther-Scharowsky-Stra{\ss}e 1, Bau 24, 91058 Erlangen,
  Germany} 
\affiliation{Department f\"{u}r Physik, Universit\"{a}t Erlangen-N\"{u}rnberg,
Staudtstra{\ss}e 7, Bau 2, 91058 Erlangen, Germany}
\affiliation{Departamento de \'Optica, Facultad de Fisica, Universidad
  Complutense, 28040 Madrid, Spain}

\author{G. Leuchs}
\affiliation{Max-Planck-Institut f\"ur die Physik des Lichts,
  G\"{u}nther-Scharowsky-Stra{\ss}e 1, Bau 24, 91058 Erlangen,
  Germany} 
\affiliation{Department f\"{u}r Physik, Universit\"{a}t Erlangen-N\"{u}rnberg,
Staudtstra{\ss}e 7, Bau 2, 91058 Erlangen, Germany}

\date{\today}

\begin{abstract}

We propose a scheme for the remote preparation of entangled matter qubits in free space. 
For this purpose, a setup of two opposing parabolic mirrors is considered, each one with a single ion trapped at its focus. 
To get the required entanglement in this extreme multimode scenario, we take advantage of the spontaneous decay, which is usually considered as an apparent nuisance. 
Using semiclassical methods, we derive an efficient photon-path representation to deal with this problem. 
We also present a thorough examination of the experimental feasibility of the scheme. 
The vulnerabilities arising in realistic implementations reduce the success probability, but leave the fidelity of the generated state unaltered. 
Our proposal thus allows for the generation of high-fidelity entangled matter qubits with high rate.

\end{abstract}


\maketitle

\section{Introduction}

The distribution of entanglement between macroscopically separated
parties constitutes a key ingredient of quantum information
networks~\cite{Nielsen,Kimble}.  A quantum network is composed of
nodes, for processing and storing quantum states, and channels linking
the nodes. The implementation of quantum nodes is a major
challenge:  different approaches are currently being pursued, most of them
involving single emitters, such as ions, atoms or nitrogen-vacancy centers~\cite{Olmschenk2,hofmann2012,Ritter,bernien2013}, even
though they are inherently probabilistic.

Photonic channels are especially advantageous, as optical photons can
carry information over long distances with almost negligible
decoherence.
In practice, there are two types of these channels:
optical fibers and free space.
Optical fibers are capable of transmitting single photons over large distances with high efficiency while suffering from effects like birefringence  or dispersion.
The free space channel, however, does not suffer from these effects, but photon losses
due to beam wandering or beam broadening, for example, can play a prominent role. Thus both types of photonic channels have their own pros and cons~\cite{Gisin} and
distribution of entangled photonic qubits was
successfully demonstrated for both of them, over a distance of 200~km~\cite{Dynes} using
optical fibers and over 144~km~\cite{Ursin07} in free space.

The main issue with a free-space channel is the low
photon-collection efficiency.  This can be improved by placing the
single emitter at the focus of a parabolic mirror~\cite{Maiwald},
which in addition enhances the atom-field interaction~\cite{leuchs2013o,
  fischer2014}. 

Here, we propose to use two opposing parabolic mirrors to
prepare maximally entangled states of two matter qubits at the
corresponding focal points. Our scheme involves an extreme multimode
scenario ,i.e., the atoms couple to a continuum of modes of the radiations field, due to the fact that the parabolic mirror is a half-open cavity. Thereby, we deal with intrinsic multimode effects like
spontaneous decay processes, which are usually considered as sources of
undesirable decoherence. Interestingly enough, we will be able to use
these effects as tools for entanglement generation, rather than
avoiding them.

In other multimode schemes~\cite{Moehring} each
deviation from the ideal situation, such as non perfect mode matching,
leads to a reduction of the fidelity of the generated state.
In contradistinction, our scheme is robust against the
vulnerabilities that arise in  experimental implementations: they 
reduce the success probability,  but leave the fidelity unaltered (and,
accordingly, it can be very high). As outlined below, this is due to the use of photons originating from
circular-dipole transitions, a suitable choice of the quantization
axis and direct dispersive probing of the qubit
states.

This paper is organized as follows. In Sec.~\ref{setup} we
advance the basic ingredients of our scheme, which is fully
analyzed in Sec.~\ref{model} by resorting to a photon-path
representation~\cite{Alber2013,Milonni} especially germane for a
multimode description. To incorporate the boundary conditions for the
relevant solution of the Helmholtz equation, we apply a semiclassical
approximation~\cite{Berry1972,Maslov}.  We discuss the results in
Sec.~\ref{entanglement} and their feasibility in
Sec. ~\ref{experiment}. Finally, our conclusions are briefly
  summarized in Sec.~\ref{V. Conclusion}.

\section{Remote entanglement preparation}
\label{setup}

\begin{figure}
  \centerline{\includegraphics[width=.90\columnwidth]{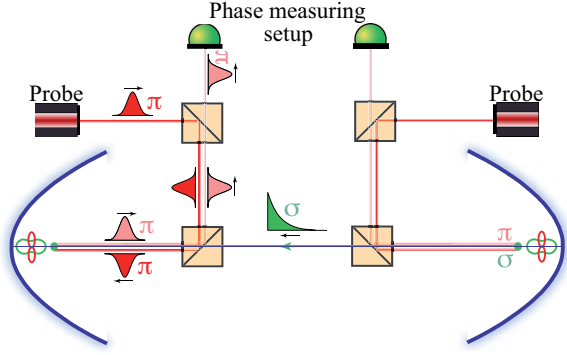}}
  \caption{(Color online) Scheme of the setup, including the post
    selection procedure: Two $^{171}$Yb$^+$ ions are trapped at the
    foci of two parabolic mirrors. Entanglement between the two ions
    is mediated by a circularly polarized photon ($\sigma$) emitted by
    ion\,1 and absorbed by ion\,2. Successful entanglement is probed
    by the dispersive interaction of weak linearly polarized coherent
    states ($\pi$) with the ions. Only if an ion resides in one of the
    desired entangled states, a phase shift is imprinted onto the
    coherent state. Probe pulses are coupled into the parabolic
    mirrors by means of beam splitters. For simplicity, the coherent
    pulses used for dispersive state detection are indicated for only
    one of the two ions.}
  \label{fig.post}
\end{figure}

Our setup, as roughly schematized in Fig.~\ref{fig.post}, consists
of two parabolic mirrors opposing each other, so they direct any
electromagnetic field from one focal point to the other with great
efficiency.

\begin{figure}
  \centering
  \includegraphics[height=0.35\columnwidth]{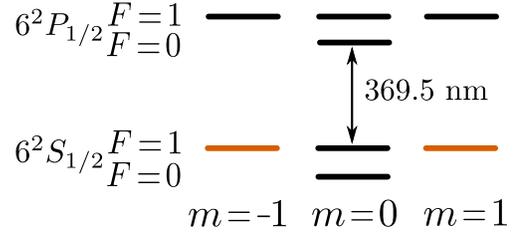}
  \caption{(Color online) Hyperfine level scheme of a
    $^{171}\text{Yb}^{+}$ ion: The states of the logical qubit,
    depicted with lighter colors, are defined by the electronic levels
    $|6^{2}S_{1/2}, F=1, m=-1\rangle$ and $|6^{2}S_{1/2}, F=1,
    m=1\rangle$.}
  \label{fig2}
\end{figure}

We consider a trapped $^{171}\text{Yb}^{+}$ ion at the focus of each
parabolic cavity.  This ion has quite a suitable hyperfine electronic
structure due to its nuclear spin $I=1/2$. We concentrate on the
level scheme formed by the levels $6^{2}S_{1/2}$ and
$6^{2}P_{1/2}$ shown in Fig.~\ref{fig2}. The logical qubit is
defined by the levels $|6^{2}S_{1/2},F=1,m=-1\rangle$ and
$|6^{2}S_{1/2},F=1,m=1\rangle$ (note the different choice
in Ref.~\cite{Olmschenk1,Olmschenk2}). The corresponding
dipole matrix elements are denoted by $ \mathbf{d}_{ij} = \langle j |
\hat{\mathbf{d}} | i \rangle$, where $|i\rangle$ and $|j\rangle$ are
the wave functions of the different states.

The basic idea is to initially prepare ions 1 and 2 in the states
\begin{eqnarray}
  |\psi^{(1)} (0) \rangle = 
  |6^{2}P_{1/2},F=1,m=0\rangle \, , \nonumber \\
  \\
  |\psi^{(2)} (0) \rangle =
  |6^{2}S_{1/2},\; F=1,m=0\rangle\, , \nonumber 
\end{eqnarray}
and use the time evolution to generate an entangled state. For that,
we notice that ion 1 can decay into three different states:
$|6^{2}S_{1/2},F=1,m=-1\rangle$, emitting a right-circularly
  polarized ($\sigma_{+}$) photon, $|6^{2}S_{1/2},F=1,m=1\rangle$,
emitting a left-circularly polarized ($\sigma_{-}$) photon, and
$|6^{2}S_{1/2}, F=0,m=0\rangle$, emitting a linearly
  polarized ($\Pi$) photon. Because we do not know which of the three
mentioned processes actually take place, the complete state of the
system is a linear superposition of the three corresponding
probability amplitudes.  As a consequence, ion 1 and the radiation
field get entangled.

The geometry of the setup ensures that the photon wave packet
generated by the spontaneous decay of ion 1 propagates to the focus of
the second parabola, where it may excite ion 2. After the absorption, the
second ion is in the state $|6^{2}P_{1/2},F=1,m=1\rangle$,
if it absorbs a $\sigma_{+}$ polarized photon, and in the state
$|6^{2}P_{1/2},F=1,m=-1\rangle$, if it absorbs a
$\sigma_{-}$ polarized photon. These absorption processes map the field state onto the state
of the second ion and thereby generating entangled matter states.
 
The ion 2 being in an excited
state (in the $6^{2}P_{1/2},\; F=1$ manifold), is affected by
spontaneous decay. So, we have to perform a state transfer from the
manifold $6^{2}P_{1/2}$ to the manifold $6^{2}S_{1/2}$ that is
radiatively stable.  One might think in using a single $\pi$-pulse,
but this is not a proper solution because the photon wave packet
radiated by ion 1 has a certain temporal width, which yields a
probabilistic determination for the time when the photon is absorbed
by ion 2. If ion 2 is still in the ground state when we apply the
$\pi$-pulse, the pulse does not have the desired effect. If we wait a
certain time to make sure that the absorption has already taken place
before applying the $\pi$-pulse, it is also likely that the
spontaneous decay process back to the $6^{2}S_{1/2}$ manifold may have
already occurred. We remind that unit excitation probability
can only be achieved with a time-reversed single-photon wave
packet~\cite{MSt}.

We suggest to use instead the spontaneous decay itself.  To that end,
we have to take into account the different decay channels.  For
example, consider that ion $2$ is in the state
$|6^{2}P_{1/2},F=1,m=1\rangle$: it can decay into the states
$|6^{2}S_{1/2},F=1,m=0\rangle$,
$|6^{2}S_{1/2},F=0,m=0\rangle$ and
$|6^{2}S_{1/2},F=1,m=1\rangle$.  But only the last process
generates entanglement.  In a similar
way, one can treat the case that ion $2$ is in the state
$|6^{2}P_{1/2},F=1,m=-1\rangle$.

\begin{figure*}
  \centering
  \includegraphics[height=0.45\columnwidth]{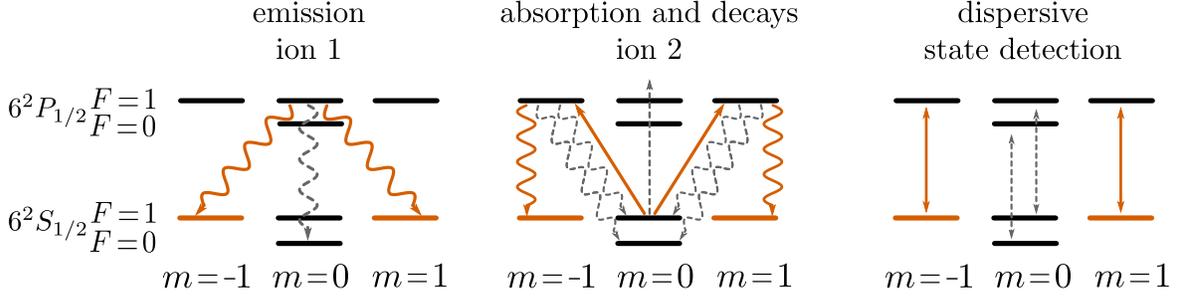}
  \caption{
    (Color online) Sequence of the processes which are the building
    blocks for the remote entanglement preparation: The states of the
    logical qubit, depicted with lighter colors, are defined by the
    electronic levels $|6^{2}S_{1/2}, F=1, m=-1\rangle$ and
    $|6^{2}S_{1/2}, F=1, m=1\rangle$. Optical transitions which are
    necessary for the entanglement generation are indicated by solid
    arrows, whereas the undesired transitions are indicated by dashed
    arrows.  The three columns correspond to three phases. The first
    column shows the possible decay channels of the first ion's
    initially prepared state $|6^{2}P_{1/2},F=1,m=0\rangle$; the
    second column shows the possible excitation procedures of the
    second ion which was initially prepared in the state
    $|6^{2}S_{1/2},F=1,m=0\rangle$ followed by the spontaneous decay
    processes used to accomplish the state transfer from the
    $6^{2}P_{1/2}$ manifold to the radiatively stable $6^{2}S_{1/2}$
    manifold; the last column shows the optical transitions used to
    perform the postselection procedure based on hyperfine splitting
    and off-resonant matter field interactions.  }
  \label{fig3}
\end{figure*}
To discard the undesired decay processes, we have to perform a
postselection. Since only in case of successful entanglement
generation both ions end up in the qubit state, by probing the
occupation of the qubit states we discard the undesired decay
processes.  This can be performed with negligible loss of entanglement
by using dispersive state detection.  It suffices to couple weak off-resonant
coherent pulses to the $\Pi$ transitions from ${S_{1/2},F=1
  ,m=\pm 1}$ to ${P_{1/2},F=1 ,m=\pm 1}$: population
is then detected by the phase shifts imprinted onto the coherent
states. This procedure allows us to check the population of the qubit
states while preserving the possible linear superpositions and thus
does not disturb the entangled state.

Furthermore, this postselection also detects photon losses, so that the
scheme is loss tolerant. This is due to the fact that upon photon loss ion\,2 remains in
$|S_{1/2},F=1,m=0\rangle$ and post selection is probed on
$\Pi$-transitions, which are for $|S_{1/2},F=1,m=0\rangle$
either forbidden or detuned so strongly that no phase shift of the
probe pulse occurs. Of course, losses reduce the success
probability, but the fidelity after a successful postselection is not
affected. The low success probability can be overcome with a high
repetition rate.

All the steps for generating entangled states described above are depicted in Fig.~\ref{fig3}.

\section{Theoretical analysis}
\label{model}

\subsection{System Hamiltonian}
\label{sub2.2}

In the rotating-wave and dipole approximations, the Hamiltonian of the
foregoing system can be written as 
\begin{equation}
  \hat{H} = \hat{H}_{A}+ \hat{H}_{R} + \hat{H}_{AR} \, ,
\end{equation}
where
\begin{eqnarray}
  \hat{H}_{A} &=& \sum_{i\in S_{e}\cup S_{g}}  \sum_{j\in S_{e}\cup   S_{g}}
\hbar (\omega_{i} + \omega_{j}) \, 
 | i^{(1)}\rangle\,\langle i^{(1)}| \otimes|j^{(2)}\rangle\,\langle j^{(2)}|, 
  \nonumber \\
  \hat{H}_{R} & = & \sum_{r} \hbar \omega_{r} \,
  \hat{a}_{r}^{\dagger} \hat{a}_{r} , \\
  \hat{H}_{AR} & = & -\sum_{\alpha\in \{ 1,2\}} 
  \hat{\mathbf{E}}^{+}(\mathbf{x}_{\alpha}) \cdot
  \hat{\mathbf{d}}_{\alpha}^{-}+
  {\rm  H.\, c.}  \nonumber 
  \label{Hamilton}
\end{eqnarray}
Here, $\hat{H}_{A}$ describes the dynamics of the matter.  We indicate
by $S_{e}$ (excited) the set of states in the manifold
$6^{2}P_{1/2}$ and by $S_{g}$ (ground) the set of states in
the manifold $6^{2}S_{1/2}$.  The vectors $|i^{(1)} \rangle$,
$|j^{(2)} \rangle$ represent states of the ion 1 and ion 2 living in
$i,j \in S_{e} \cup S_{g}$, with energies $\hbar \omega_{i}$ and
$\hbar \omega_{j}$, respectively.  $\hat{H}_{R}$ gives the dynamics of
the field, characterized by the annihilation ($\hat{a}_{r}$) and
creation ($\hat{a}_{r}^{\dagger}$) operators of the modes (of
frequency $\omega_{r}$) that couple to the ions (they depend on the
boundary conditions).  Finally, the interaction between the ions and
the field is given by $\hat{H}_{AR}$, wherein H.c. stands for the
Hermitian conjugate and
\begin{eqnarray}
  \hat{\mathbf{E}}^{+} (\mathbf{r} ) & = & - i   
  \sum_{r}\sqrt{\frac{\hbar\omega_{r}}{2\epsilon_{0}}} 
  \mathbf{g}_{r} ( \mathbf{r} ) \, \hat{a}_{r}^{\dagger} , \nonumber \\
  & &\\
  \hat{\mathbf{d}}_{\alpha}^{-} & = & \sum_{i \in S_{e}}  
  \sum_{j \in S_{g}} \mathbf{d}_{ij} \, 
  |j ^{(\alpha)} \rangle \, \langle i^{(\alpha)} |  , \nonumber 
\end{eqnarray}
$\mathbf{x}_{1}$ and $\mathbf{x}_{2}$ being the position of the first
and second ion, respectively.  The orthonormal mode functions
$\mathbf{g}_{r} (\mathbf{r})$ are solutions of the Helmholtz equation
with the proper boundary conditions, fulfilling, in addition, the
transversality condition $\nabla \cdot \mathbf{g}_{r} (\mathbf{r}
)=0$.

\subsection{Photon-path-representation}
\label{sub2.3}

Since only one excitation is available in our initial state, and the
Hamiltonian \eqref{Hamilton} preserves the number of excitations, the
state of the system at time $t$ can be written as
\begin{eqnarray}
  \label{eq:ans}
  | \psi(t) \rangle  & = & \sum_{i \in S_{e}} 
  \sum_{j \in S_{g}}  b_{ij}^{(1)} (t)  \,  |i^{(1)} \rangle
  | j^{(2)} \rangle | \{ 0 \} \rangle \nonumber \\
  & + & \sum_{i \in S_{g}}  \sum_{j\in S_{e}}  
  b_{ji}^{(2)}(t) \, | i^{(1)} \rangle |j^{(2)} \rangle 
  | \{ 0 \}  \rangle \nonumber \\
  & + & \sum_{r} \sum_{i \in S_{g}} 
  \sum_{i \in S_{g}}  f_{ij}^{(r)} (t)  \, | i^{(1)} \rangle
  | j^{(2)} \rangle | 1_{r} \rangle  ,
\end{eqnarray}
where $| \{0 \} \rangle$ is the vacuum state 
and $ | 1_{r} \rangle = \hat{a}^{\dagger}_{r} | \{ 0 \} \rangle$ a
single-photon state of the radiation field.  The
amplitude $b_{ij}^{(1)}(t)$ describes the evolution when the field is
in the vacuum, the first ion is in one of the excited levels $i\in
S_{e}$ and the second ion is in one of the ground levels $j\in S_{g}$
and an analogous interpretation for $b_{ji}^{(2)}(t)$.  The amplitude
$f_{ij}^{(r)}(t)$ is related to the evolution when there is an
excitation in the field mode $r$ and both ions are in one of the
ground electronic levels $i,j\in S_{g}$.

Now, we can solve the time-dependent Schr\"{o}dinger equation, with
the \emph{ansatz} \eqref{eq:ans}.  If we assume that the 
field is initially in a vacuum state and we use the Laplace transform, we
get, after eliminating the transforms of the probability amplitudes
for photonic excitations $\widetilde{f}_{ij}^{(r)}(s)$, 
\begin{align}
 s & \, \widetilde{b}_{ij}^{(\alpha)} (s) - b_{ij}^{(\alpha)} (0) = 
- i (\omega_{i} + \omega_{j} ) \, \widetilde{b}_{ij}^{(\alpha)} (s) \nonumber \\
 + &    \sum_{\beta \in \{ 1,2 \} }
 \sum_{k \in S_{e}} \sum_{\ell \in S_{g}} 
T_{\beta;k\ell}^{\alpha;ij} (s) \, \widetilde{b}_{k\ell}^{(\beta)}(s) \, ,
\label{eq:compact_scalar_equation-laplace_transform}
\end{align}
where $ \alpha \in \{ 1,2 \}$ indexes the ions, $i\in S_{e},\; j\in S_{g}$ and 
\begin{equation}
T_{\beta;k\ell}^{\alpha;ij} (s ) = 
\left \{ 
\begin{array}{ll}
\displaystyle 
\delta_{j \ell} \sum_{m \in S_{g}} 
 A_{\alpha;im}^{\beta;km} (s + i \omega_{m}+i \omega_{j}  ), & \alpha =\beta,\\
A_{\alpha;i\ell}^{\beta;kj} (s+i \omega_{j}+i \omega_{\ell} ), &
\alpha\neq \beta.
\end{array}
\right .
\end{equation}
The function
\begin{equation}
A_{\beta; k\ell}^{\alpha;ij} (s ) 
=  \sum_{r}\frac{\kappa_{r}^{\alpha;ij}
  \left(\kappa_{r}^{\beta; k\ell}\right)^{\ast}}{s+i \omega_{r}} ,
\end{equation}
with 
\begin{equation}
\kappa_{r}^{\alpha;ij} =
\sqrt{\frac{\omega_{r}}{2\epsilon_{0}\hbar}}
\mathbf{d}_{ij}^{\dagger}  \cdot \mathbf{g}_{r}
  (\mathbf{x}_{\alpha} )  \, ,
\end{equation}
describes all possible photon emission and absorption processes and
encodes the whole geometry of the setup through the modes
$\mathbf{g}_{r}(\mathbf{r})$.  Its explicit calculation turns to be a
difficult task, although in Appendix~\ref{sub2.4} we sketch a
semiclassical method.

Equation~\eqref{eq:compact_scalar_equation-laplace_transform} can be
recast in a suggestive vectorial form
\begin{equation}
  (s+i \omega+ T) \, \widetilde{\mathbf{b}} (s) =\mathbf{b}(0)\;,
  \label{Tsystem}
\end{equation}
where the functions $\widetilde{b}_{ij}^{(a)}$ are arranged in a
vector $\widetilde{\mathbf{b}}$ and $i \omega$ is a diagonal matrix
which represents the term $i (\omega_{1} +\omega_{2} )$. The
contribution $T$ can be split as
\begin{equation}
  T=T_{0}+T_{1}\;,
\end{equation}
where $T_{0}$ is equal to $\frac{3}{2}\Gamma$, with $\Gamma$ being the
spontaneous decay rate in free space, and $T_{1}$ embodies
exponentially decaying terms of the form $e^{-s t}$. In principle, one
could try to perform an inverse Laplace transform to
solve~\eqref{Tsystem}. However, this involves finding the poles of the
integrand, which is a formidable task because $T$ depends itself on
$s$ in a highly nontrivial way.

To determine $\widetilde{\mathbf{b}}(s)$ we use instead an alternative
route based on the Neumann expansion
\begin{eqnarray}
  (\openone -K)^{-1} = \openone + K + K^{2} +\dots+ K^{n} +\dots 
\end{eqnarray}
If we take $K:=-T_{1}(s+i \omega+\frac{3}{2}\Gamma)^{-1}$ we get
\begin{equation}
  \widetilde{\mathbf{b}} (s) = 
  \sum_{n=0}^{\infty}  (s+i \omega+ 3\Gamma /2 )^{-1} K^{n} 
  \mathbf{b} (0)  \, ,
  \label{eq:Neumann_expansion}
\end{equation}
and each term in the series can be immediately Laplace inverted. The
price we pay is that we have to deal with an infinite series.

In most circumstances, only a few terms contribute to
  Eq.~\eqref{eq:Neumann_expansion}: since 
 each summand  is damped by an  exponential of the form $e^{-s t}$, and
when applying the inverse Laplace transform, each term
leads to a Heaviside step function, i.e., a retardation, shifted into
the positive direction by time $t$. If we are looking at the evolution
in a finite time interval, we can neglect terms which are so far
retarded that they do not contribute.

The time interval of interest in our setup is of the order of $\tau =
(4f+d)/c$,  which is the typical travel time of a photon to go
from the first to the second ion and $f$ being the focal length of the mirrors and $d$ the
separation between foci. In consequence, we can neglect all terms
$n>1$ in the sum; as shown in Ref.~\cite{Alber2013}, the higher-order
terms are relevant when the focal length is comparable with the
wavelength, which is not the case for our actual mirrors~\cite{Maiwald}
($f = $2.1 mm and wavelength $\lambda =369$ nm). 

\section{Results}
\label{entanglement}

If one uses the method of the preceding section in the time interval
$t \in[0, 2\tau)$, it turns out that only four of the atomic
probability amplitudes $b_{ij}^{(\alpha)} $ ($ \alpha \in \{1, 2\} $) are of
relevance:
\begin{equation}
  \begin{array}{c}
    b_{6^{2}P_{1/2},\; F=1\; m=0,6^{2}S_{1/2},\; F=1\; m=0}^{(1)}(t),\\
    b_{6^{2}P_{1/2},\; F=1\; m=1,6^{2}S_{1/2},\; F=1\; m=-1}^{(2)}(t),\\
    b_{6^{2}P_{1/2},\; F=1\; m=-1,6^{2}S_{1/2},\; F=1\; m=1}^{(2)}(t),\\
    b_{6^{2}P_{1/2},\; F=0\; m=0,6^{2}S_{1/2},\; F=0\; m=0}^{(2)}(t).
  \end{array}
\end{equation}
If the hyperfine splitting is large in comparison to the spontaneous
decay rate $\Gamma$, as it happens for $^{171}\text{Yb}^{+}$ in the
time window of interest, we can also neglect
$b_{6^{2}P_{1/2},\; F=0\; m=0,6^{2}S_{1/2},\;
  F=0\; m=0}^{(2)}(t)$.

\begin{figure}[t]
  \begin{centering}
    \includegraphics[width=0.95\columnwidth]{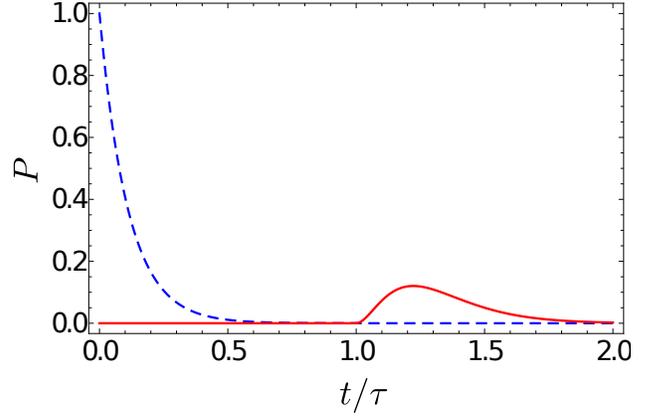}
  \end{centering}
  \caption{(Color online) Time evolution of the excitation probability
    $P$ in the case of ion\, 1 (dashed line) and of ion\, 2 (solid
    line): The interaction time $t$ is plotted in units of the time
    $\tau=(4f+d)/c$ which a photon needs to travel from the first ion
    to the second ion.  $f$ is the focal length of the parabolas and
    $d$ is the distance between the foci. We set the spontaneous decay
    $\Gamma\tau=3$ and the Zeeman splitting was neglected.}
  \label{fig6}
\end{figure}

In Fig.~\ref{fig6} we plot the excitation probabilities of the two
ions for vanishing Zeeman splitting.  As discussed in
Sec.~\ref{setup}, the  process generates an entangled state if one uses the states
$|6^{2}P_{1/2},\; F=1\; m=1\rangle$ and
$|6^{2}P_{1/2},\; F=1\; m=-1\rangle$ of the second ion as
qubit. But these states do not form a stable qubit; spontaneous decay
transfers them to the ground states $|6^{2}S_{1/2},\; F=1\;
m=1\rangle$ and $|6^{2}S_{1/2},\; F=1\; m=-1\rangle$
by emitting a single photon.  By detecting whether or not both ions
are in one of the states $|6^{2}S_{1/2},\; F=1\;
m=\pm1\rangle$ , we check if the entanglement generation was
successful.  

The postselection is equivalent to a von Neumann measurement described
by the projection operator
\begin{equation}
  \hat{P}=\ket{00}\bra{00}+\ket{01}\bra{01}+
\ket{10}\bra{10}+\ket{11}\bra{11},
\end{equation}
where $\ket{q_{1}q_{2}}$ ($q_1,q_2\in\left\{ 0,1\right\}$) correspond
to the states of the logical qubit.  In addition, we also have to deal
with the photon emitted in the transfer from the excited to the ground
states.  This photon, which carries information about the 
state of the ions, might cause decoherence and therefore destroy the
entangled state generated by the time evolution. To certify that this
is not the case, we have to trace out the uncontrolled photonic
degrees of freedom, which amounts to know
\begin{equation}
  \hat{\varrho} (t)  =  
\Tr_{R} ( | \psi(t) \rangle \langle \psi(t) |  )   \, .
\end{equation}
This density matrix is evaluated in Appendix ~\ref{subTracing_out}. In the limit $\Gamma (t-\tau)\rightarrow\infty$ with
$\tau<t<2\tau$, we have that
\begin{small}
  \begin{align}
    \hat{P} & \hat{\varrho}(t) \hat{P} = \frac{2}{3 (9+\delta^{2})} \,
    | 01 \rangle \langle 01| + \frac{2}{3\left(9+\delta^{2}\right)} |
    10 \rangle \langle 10|
    \nonumber \\
    + & \frac{2}{3[-9+\delta(-9i+2\delta)]} | 01 \rangle \langle 10| +
    \frac{2}{3[ -9+\delta(9i+2\delta)]} | 10 \rangle \langle 01| \; .
  \end{align}
\end{small}
The parameter $\delta= ( \Delta_{1}-\Delta_{2})/\Gamma$ characterizes
the Zeeman splitting of the energy levels: $\Delta_{1} m$ is the
splitting in the $6^{2}P_{1/2},\; F=1$ manifold and
$\Delta_{2} m$ the splitting in the $6^{2}S_{1/2},\; F=1$
manifold. Note that the magnetic field has the same orientation and
strength for both ions.

For $| \delta | \ll1$, which is justified in our experimental
setup~\cite{Maiwald}, we get
\begin{equation}
  \hat{P}\hat{\varrho}(t)\hat{P} = 
  \frac{2}{27}  ( | 01 \rangle-| 10 \rangle) (\langle  01|-\langle  10|)  \, ,
\end{equation}
which corresponds to a maximal entangled state with a success
probability $4/27 \approx 15~\%$. Of course, in a real experiment, one
has to take additional effects into account. As we explore in
the next Section, it should be possible to achieve free-space
communication over several kilometers.

\section{Experimental feasibility}
\label{experiment}

\subsection{Realistic parabolic mirrors}
 \label{sub:Non_perfect_boundary_conditions}

 In a real setup, the parabolic mirror does not cover the full solid
 angle. Actually,  in our  parabolic
mirror~\cite{Maiwald} we have
 \begin{equation}
   \Omega = \left\{ (\varphi,\theta) \, : \,
     \varphi\in ( 0, 360^{\circ} ) , 
     \theta \in (20^{\circ}, 135^{\circ} ) \right\} \, ,
 \end{equation}
 whereby the angle $135^{\circ}$ gives the front opening of the
 parabola and the angle $20^{\circ}$ accounts for the  hole on
 the backside for inserting the ion trap. This has to be taken into
 account in integrations as in Eq.~\eqref{eq:Gamma_rel}.

 Furthermore, our mirrors are made out of aluminum, which has a finite
 electrical conductivity.  The properties of the material are well
 described by introducing a frequency dependent dielectric constant
 $\epsilon(\omega)$. In our case
 $\epsilon(\omega)=-18.74+i3.37$~\cite{refractiveindex}.  Now, we have
 to split the field in a transverse electric (TE) and a magnetic (TM)
 part and apply Fresnel equations to deal with the boundary
 conditions. But these equations are different for the two basic
 polarizations and give angle-dependent phase shifts and
 reflectivities, which leads to a further reduction of the efficiency
 for entanglement generation.  One might think that this effect could
 also reduce the fidelity of the entangled state but this is not the
 case.  Such a reduction could occur if a $\sigma_{+}$($\sigma_{-}$)
 decay of the first ion could drive a $\sigma_{-}$($\sigma_{+})$
 transition of the second ion, but due to the symmetry
 this does not occur.
 
This is obvious from the following reasoning: After collimation by the
parabolic mirror, the polarization vector of the electric field in the
exit pupil of the parabolic mirror reads~\cite{sondermann2008}
\begin{align}
\label{eq.psigma}
\mathbf{\sigma}_\pm \simeq&   
(r^2 - 4)(\cos{\phi}\pm i\sin{\phi})\cdot \mathbf{e}_r\\
& + (r^2 + 4)(\sin{\phi}\mp i\cos{\phi})\cdot \mathbf{e}_\phi \nonumber
\end{align}
with $r$ the distance to the optical axis in units of the mirror's
focal length, $\phi$ the azimuthal angle, and $\mathbf{e}_r$ and
$\mathbf{e}_\phi$ the unit vectors in radial and azimuthal
direction, respectively.
Upon reflection on the parabolic surface, these vectors correspond to
TM- and TE-components. 
The influence of the metallic mirror can be accounted for by
additional complex pre-factors which depend on $r$ only.  
It is straightforward to show that the overlap $\int
\tilde{\mathbf{\sigma}}_\pm\cdot\mathbf{\sigma}^\star_\mp$ of this
modified field $\tilde{\mathbf{\sigma}}_\pm$ with the state of
opposite helicity vanishes.   

We can sum all the above effects in a factor $\eta$ which has to be
multiplied with the probability for a successful entanglement
creation to take the more realistic mirrors into account: $\eta=1$
corresponds to perfectly conducting parabolic mirrors that cover the
full solid angle.  In the specific case treated here, we have
$\eta\approx 0.47$.

\subsection{Free-space versus fiber-based transmission}
\label{Freespace_vs_fiber}

Our scheme is designed to be compatible with free-space
communication by photonic qubits, for it does not rely on the
  strong coupling regime, but on intrinsic multimode effects like
  spontaneous emission.

There are other multimode schemes, such as the
  one in Ref.~\cite{Moehring}, which might be adapted to free-space
communication, but our proposal offers  considerable advantages. 
The scheme in Ref.~\cite{Moehring} heavily relies on  fibers as mode filters
to achieve almost perfect mode matching on a beam splitter and,
besides,  the fidelity  is mainly limited by the fact that the postselection
is performed on the radiation field and is sensitive to dark counts of
the detectors. In contrast,  in our proposal, postselection is
performed on the ions, which circumvents detector dark
counts.

Of course, we have to take into account experimental
  imperfections, mainly connected with the free-space transmission of
  the one-photon wave packet.  This gives rise to beam wandering and
phase-front distortions due to atmospheric
turbulences~\cite{Ursin07}. In both cases, the intensity at the focus
is reduced~\cite{lieb2001,leuchs2008,april2011}, affecting the success
probability.  Once the distance between the two parabolic mirrors
becomes large enough, beam broadening plays a crucial role, which also
results in a lower success probability. All these effects,
  however, diminish the success probability but seem to leave the
fidelity rather untouched, which is of big importance for practical
applications.

One could also think about the transmission of the photon from ion 1
to ion~2 via an optical fiber.  This would circumvent all problems
related to atmospheric transmission, but, due to their complex
polarization pattern, cf. Eq.~\eqref{eq.psigma}, the photons collimated by the parabolic mirror
have subunit overlap with a fundamental Gaussian mode with circular
polarization. Hence the efficiency in coupling these photons to a
single-mode fiber is limited to a maximum of
49~\%~\cite{sondermann2008}.  Therefore, fiber transmission alone would
limit the success probability 
of our entanglement scheme to 24~\%. Moreover, the strong
attenuation of ultraviolet radiation in standard optical fibers
reduces the success probability by orders of magnitude, even for
distances about 1~km. Finally, fibers are not well suited to
perform communication via polarization coding \cite{Gisin}, as  in
our scheme.

\subsection{Postselection}
\label{Post_selection}

As advanced in Sec.~\ref{setup}, the best way to perform postselection
seems to probe qubit states directly by dispersive state
detection.  This can be implemented by coupling weak coherent-state
pulses to the $\Pi$ transitions from $S_{1/2},F=1 ,m=\pm
1$ to ${P_{1/2},F=1 ,m=\pm 1}$. Population in the
$S_{1/2},m=\pm 1$ states is then detected by the phase shifts
imprinted onto the coherent states. The detuning and pulse amplitudes
can be chosen such that one is far from saturating the respective
transitions.  For example, choosing an on-resonance saturation
parameter of $s_{0}=0.01$ and a detuning of two linewidths, the
excitation probability is about $10^{-5}$, while the phase of the
coherent pulse is shifted by $25^{\circ}$,
according to the formalism of Ref.~\cite{sondermann2013p}.

One has to balance the amplitude and the detuning of the incident coherent
state carefully.  Larger amplitudes and smaller detunings result in
lower error probabilities for detecting the phase of the coherent
state, but also enforce a stronger excitation of the ion.  The latter
might lead to transferring the ion out of the $m=\pm 1$ state during
state detection, hindering the phase shift of the coherent state and
hence resulting in erroneous postselection.  Furthermore, the
reflectivity of the beam splitters in front of the parabolic mirrors
not only affects the success probability of our entangling scheme, but
also influences the error in measuring the phase of the coherent
state.

\begin{figure}
\centerline{\includegraphics{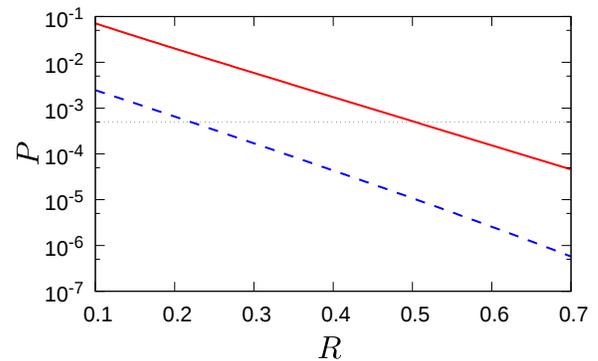}}
    
  \caption{\label{fig.detect} (Color online) Error probability $P$ in
    determining the phase of a coherent pulse probing the
    $\Pi$-transitions from $S_{1/2}$ to $P_{1/2}$ plotted over the beam splitter reflectivity $R$.  Solid line: ion~1,
    dashed line: ion~2.  In both cases, the relative detuning respect
    to the resonance is two linewidths, corresponding to a phase shift
    of $0.14\pi$.  The length of the pulse is $10^4$ upper-state
    lifetimes or 81~$\mu$s, respectively. The amplitude of the
    coherent state incident onto the ion is chosen such that the
    probability to excite the respective upper state is $5\times 10^{-4}$, as marked by the
    thin dotted line.  The calculation for ion~2 accounts for the
    threshold reflectivity found for ion~1, which is marked by the
    crossing of the solid and the dotted line.}
\end{figure}

We compute the corresponding error probabilities according to the
Helstrom bound~\cite{bergou2010}.  The \emph{a priori} probabilities
in this calculation are obtained from all relevant branching ratios,
excitation probabilities, and reflectivities.  The amplitude of the
coherent state is chosen such that the probability to excite the ions
with the probe pulse is $5\times 10^{-4}$.  We also choose this value
as an upper bound for the acceptable error probability.  This is
motivated by the fact that postselection schemes probing the
$m=0$ states are limited in fidelity to values $\le 0.995$ by the
branching ratio of the $P_{1/2}$ state to the $D_{3/2}$ state of
0.5\%.  Keeping all errors in our postselection scheme an order of
magnitude below this value is reasonable and desirable.

The minimum error probabilities as a function of the reflectivity of
the beam splitters coupling the coherent states into the parabolic
mirrors is plotted in Fig.~\ref{fig.detect}.  First, we
determined the reflectivity for the beam splitter in front of ion~1
that ensures being below the error threshold for a set of suitable
parameters, yielding $R_1=0.5$.  Next, this result was used in the
calculations for ion~2, leading to $R_2=0.22$.  From these
reflectivity values one would obtain a reduction of the success
probability for entanglement generation by 61~\%.

In practice the Helstrom bound will not be reached entirely, with the
actually obtainable error probability depending on the method employed
for measuring the phase of the probe pulse.  Nevertheless the error
threshold marked in Fig.~\ref{fig.detect} can be reached.  This may be
achieved at the cost of using beam splitters with larger
reflectivities and thus accepting lower success probabilities.

To guarantee that the entangled state is not destroyed, we have to
ensure that no information about the state of the qubit is extracted
by our postselection. The latter condition is fulfilled if the
magnetic field fixing the quantization axis is sufficiently small (,i.e., the frequency shifts caused by the Zeeman effect are small compared to the spontaneous decay rate), so that the
phase shift imprinted by an ion in the $m=-1$ Zeeman state will be
practically the same as for the other ion in the $m=+1$ state.
Therefore, probing the qubit dispersively will not project the ions
into one of these states and entanglement is preserved.  The
parameter set in Fig.~\ref{fig.detect} yields a fidelity of 0.998
when postselecting.  Even higher fidelities can be reached by larger
beam splitter reflectivities (accompanied by decreasing success
probabilities), lower pulse amplitudes or longer pulse lengths.  A
lower pulse amplitude has to be compensated for by larger beam
splitter reflectivities or longer pulse lengths.  The latter in turn
affects the repetition rate.

\subsection{Repetition rate}

We finally estimate the achievable repetition rate.  Typically, an
experimental cycle starts with Doppler cooling the ion, which takes
about 200~$\mu$s for the ions treated
here~\cite{PhysRevA.80.022502}.  For the trap frequencies inherent to
the parabolic mirror trap, 500\,kHz in radial direction and 1\,MHz
along the optical axis, the average number of motional quanta
according to the Doppler limit is 20 and 10, respectively.  This
corresponds to widths of the ion wave function in position space about
0.13 and 0.07 wavelengths.  With these numbers we estimate that the
ions experience 78~\% of the focal intensity obtained by diffraction
limited focusing.  Applying only Doppler cooling the success
probability of our entanglement scheme would be reduced accordingly.
One could additionally apply resolved side-band cooling, but the
increase of the success rate is obviously accompanied by a lower
repetition rate due to the elongated cooling procedure.  Furthermore,
as soon as there is a broadened focus due to incompletely compensated
atmospheric aberrations etc. the above spread of the ion's wave
function is negligible.

After cooling, both ions have to be prepared in the state
${S_{1/2},F=0}$ which takes less than 1~$\mu$s~\cite{Olmschenk1}.
Additionally, ion~2 has to be flipped to the state
${S_{1/2},F=1,m=0}$.  This can be accomplished in 6~$\mu$s using
microwaves~\cite{Olmschenk1} or in 100~ps applying Raman
transitions~\cite{campbell2010}.  Likewise, ion~1 is brought to the
${P_{1/2},F=1,m =0}$ state by an optical $\pi$-pulse on a time
scale smaller than a nanosecond.  The postselection
requires around 80~$\mu$s, as it was outlined in
Sec.~\ref{Post_selection}.  The photon traveling time  from
ion~1 to ion~2 is of the order of 10~$\mu$s for distances of a few
kilometers.  At least, the same time has to be spent in communicating
the postselection via a classical communication channel.  Thus,
the time spent for state preparation, attempting entanglement of the
ions and postselection is on the order of 100\,$\mu$s.

From the numbers given above, one would estimate a repetition rate of
3.3\,kHz if Doppler cooling is applied after each entanglement
attempt.  One could increase the repetition rate if Doppler cooling is
performed regularly after a certain number of entanglement trials.
Since one entanglement trial takes about 100\,$\mu$s, a repetition
rate in excess of 10\,kHz is not feasible, unless one accepts a
reduced fidelity and/or success probability.  Assuming a realistic
heating rate of 10 quanta per ms~\cite{mcloughlin2011}, the
spread of the ion wave function would roughly double in radial
direction within 8~ms.  Accepting the accompanying, continuously
increasing loss of success probability, one could enhance the
repetition rate towards 9.8~kHz, which is close to the inverse of the
duration of one entanglement trial.  Anyhow, in every
  experimental realization,  the repetition rate is dictated by the
specific requirements on fidelity, success probability and inter-ion
distance.

\section{Concluding remarks}
\label{V. Conclusion}

In summary, we have presented a scheme for preparing maximally
entangled states of two matter qubits with high fidelity by using a
free-space channel. The  qubits are encoded in the level
structure of two distant $^{171}\text{Yb}^{+}$ ions located
at the foci of two parabolic mirrors.  The theoretical
description of the setup involves an extreme multimode
scenario to model the radiation field and  a level structure
far more complicated than a simple two level atom. 

We have used a semiclassical photon-path representation to deal
with the boundary conditions at the two parabolic mirrors, which leads
to a intuitive representation of the quantum dynamics of the two ions
and the radiation field.

To obtain a more realistic description, we have focused on the
experimental details in Ref.~\cite{Maiwald} and on more
  realistic boundary conditions. Our results confirm the feasibility
of the scheme to achieve reasonable success probabilities, which in
combination with a relatively high repetition rate leads to a proper
rate for preparing entangled matter qubits. Indeed, we expect an
  entanglement rate of 54 per second under diffraction-limited
  focusing.

One of the main issues is the fidelity of these states. Our scheme is
robust against imperfections arising in the experimental
implementation. All these effects reduce the success
probability of entanglement generation, but leave the fidelity 
untouched.

We hope that our work is a step towards an experimental realization of
remote entangled matter qubits in free space, which is a key building block for
future quantum technologies.

\begin{acknowledgments}
  N.~T., J.~Z.~B., and G.~A.  acknowledge support by the BMBF
  Project Q.com and CASED III. M.~S. and G.~L. are grateful for the
  financial support of the European Research Council under the
  Advanced Grant PACART. Finally, L.~L.~S.~S. acknowledges   the
  support from the Spanish MINECO (Grant No. FIS2011-26786). 
\end{acknowledgments}

\appendix

\section{Determination of the functions  $A_{\beta;k\ell}^{\alpha;ij}$}
\label{sub2.4}

We describe here how to evaluate the functions
$A_{\beta;k\ell}^{\alpha;ij} (s)$.  The main idea is to relate these
functions with the Laplace transform of the retarded Green's functions
of the vectorial d'Alembert operator, which can be determined by using
the multidimensional JWKB approximation. This is valid when the
typical wavelength $\lambda$ is much smaller compared to the focal
length of the cavities $f$. Besides, this will enable us to clarify
the retardation effects in $A_{\beta;k\ell}^{\alpha;ij} (s)$ due to
the propagation of a photon wave packet.

Let us introduce the functions
\begin{equation}
  B_{\beta;k \ell}^{\alpha;ij} (s )=-\frac{i}{\epsilon_{0}\hbar} 
  \mathbf{d}_{ij}^{\dagger} \, \mathcal{L} [\nabla\times\nabla\times 
  G(\mathbf{x}_{\alpha},\mathbf{x}_{\beta},t)] \, \mathbf{d}_{k\ell}
\end{equation}
where $\mathcal{L}$ denotes the Laplace transform and $G$ is the
Green's function of the vectorial d'Alembert operator satisfying the
appropriate boundary conditions. We recall that $G$ can be
expanded in terms of the mode functions as
\begin{equation}
  G (\mathbf{x},\mathbf{x}^{\prime},t) = c^{2} \sum_{r}
  \mathbf{g}_{r}(\mathbf{x}) \otimes \mathbf{g}_{r}(\mathbf{x}^{\prime})
  \frac{\sin (\omega_{r}t)}{\omega_{r}} \Theta (t) \, ,
\end{equation}
where $\otimes$ is  the dyadic product and $\Theta (t)$ the Heaviside
step function. We can immediately show that  
\begin{eqnarray}
B_{\beta;k\ell}^{\alpha;ij} (s ) & = &  \frac{1}{2\epsilon_{0}\hbar} 
\sum_{r} \mathbf{d}_{ij}^{\dagger} 
\mathbf{g}_{r}(\mathbf{x}_{\alpha}) \otimes 
\mathbf{g}_{r}(\mathbf{x}_{\beta})
\mathbf{d}_{k\ell}\nonumber \\
& \times & \left ( \frac{\omega_{r}}{s+i \omega_{r}}
 -\frac{\omega_{r}}{s-i \omega_{r}} \right ) \, .
\end{eqnarray}
If we compare with our definition of $A_{\beta;k\ell}^{\alpha;ij}
(s)$, viz
\begin{equation}
A_{\beta;k\ell}^{\alpha;ij} (s )=
\frac{1}{2\epsilon_{0}\hbar} \sum_{r} 
\frac{\omega_{r}}{s+i \omega_{r}}
 \mathbf{d}_{ij}^{\dagger}  \, \mathbf{g}_{r}(\mathbf{x}_{\alpha})
 \otimes\mathbf{g}_{r}(\mathbf{x}_{\beta}) \, 
\mathbf{d}_{k\ell} \, ,
\end{equation}
we see that the two expressions just differ by
\begin{equation}
A_{\beta;k\ell}^{\alpha;ij} ( s )  -  B_{\beta;k\ell}^{\alpha;ij} (s )=
\frac{1}{2\epsilon_{0}\hbar} \sum_{r} 
\frac{\omega_{r}}{s-i \omega_{r}}  \mathbf{d}_{ij}^{\dagger} \, 
\mathbf{g}_{r}(\mathbf{x}_{\alpha})\otimes
 \mathbf{g}_{r}(\mathbf{x}_{\beta}) \, 
\mathbf{d}_{k\ell} \; .
\end{equation}
This term can be neglected by using the same argument employed to
justify the rotating-wave approximation and,  therefore,  to a good
approximation, we can identify $A_{\beta;k\ell}^{\alpha;ij} (s )$
with $B_{\beta;k\ell}^{\alpha;ij} (s )$.

Our next step is to get a manageable expression for these functions.
To achieve this we  divide our cavity in three regions:
\begin{enumerate}
\item A sphere of radius $R$ centered around the first ion.
$R$ has to be chosen such that $R \gg\lambda$, but small when compared
with  $f$ and $d$. 
\item The whole volume,  except the  spheres centered around
the ions. 
\item A sphere of radius $R$ centered around the second ion. 
\end{enumerate}
In regions 1 and 3, we use the free-space propagator, while in region
2 we use a  JWKB approximation for the Green's function (which is
presented in Appendix~\ref{sub:Multidimensional JWKB-Method}).
After matching the resulting expressions, we obtain 
\begin{eqnarray}
\label{eq:resfin}
A_{1;k\ell}^{1;ij} (s ) & = &  A_{2;k\ell}^{2:ij} (s ) = 
\frac{1}{2} \mathbf{d}_{ij}^{\dagger} \mathbf{d}_{k\ell} 
\frac{\omega^{3}}{3\pi c^{3}\epsilon_{0}\hbar} ,  \nonumber \\
& & \\
A_{2;k\ell}^{1;ij} (s ) & = &  A_{1;k\ell}^{2:ij} (s ) =
- \mathbf{d}_{ij}^{\dagger} \Gamma_{\text{rel}} \mathbf{d}_{kl} \, 
 \frac{\omega^{3}}{3\pi c^{3}\epsilon_{0}\hbar} 
 \,  e^{-\tau s} \, , 
\nonumber 
\end{eqnarray}
where 
\begin{eqnarray}
\label{eq:Gamma_rel}
\Gamma_{\text{rel}} & = & \frac{3}{8\pi} \int_{\Omega}
 \sin\theta \, P_{\perp\mathbf{e}_{r_2}} d \theta_2d\varphi_2\; , 
 \nonumber \\
& & \\
P_{\perp\mathbf{e}_{r_{2}}} & = & \openone
-\mathbf{e}_{r_{2}}\otimes\mathbf{e}_{r_{2}} \; ,   \nonumber 
\end{eqnarray}

$\Omega$ denotes the solid angle around
the ions covered by the parabolic mirrors and $\omega$ denotes
the transition frequency of the corresponding optical transitions.

\section{The multidimensional JWKB method}
\label{sub:Multidimensional JWKB-Method}

As heralded in appendix~\ref{sub2.4}, we derive here semiclassical
approximations for the functions $A_{\beta;k\ell}^{\alpha;ij}$ by
using multidimensional JWKB method~\cite{Maslov}. In region 1,
  we use the free-space Green's function
\begin{equation}
  G_{\text{free}} (\mathbf{x},\mathbf{x}^{\prime}, t ) = 
  \frac{1}{4\pi | \mathbf{x} -\mathbf{x}^{\prime}|}  
  \delta ( t  - | \mathbf{x} - \mathbf{x}^{\prime} |/c) \; ,
\end{equation}
and, since $\lambda \ll R$, we can use the
approximation
\begin{eqnarray}
\nabla & \times &  \nabla \times 
G_{\text{free}} (\mathbf{x},\mathbf{x}^{\prime}, t ) \simeq 
-\frac{\delta^{\prime \prime} (t - |\mathbf{x} -
  \mathbf{x}^{\prime}|/c)}
{4\pi  c^{2} | \mathbf{x} -\mathbf{x}^{\prime} |} \nonumber \\
& \times & \left ( \openone -
 \frac{\mathbf{x}-\mathbf{x}^{\prime}}
 {| \mathbf{x}-\mathbf{x}^{\prime} |} \otimes
\frac{\mathbf{x}-\mathbf{x}^{\prime}}
 {| \mathbf{x}-\mathbf{x}^{\prime} |} \right ) \;.
\end{eqnarray}
If we introduce for each focus a system of spherical coordinates, with
the focus lying at the origin, we can represent this Green's function in region 1 as
\begin{equation}
\nabla \times \nabla \times 
G_{\text{free}}(\mathbf{x}_{1},r_{1},\theta_{1},\varphi_{1}, t ) = 
-\frac{\delta^{\prime \prime} ( t- r_{1}/c)}{4\pi c^{2} r_{1}} \, 
P_{\perp\mathbf{e}_{r_{1}}}\;,
\end{equation}
where $P_{\perp\mathbf{e}_{r_{1}}}$ has been defined in Eq.~\eqref{eq:Gamma_rel}.
                                          
We use the multidimensional JWKB method to propagate this
  expression to the second focus; that is, in region 2. Therefore, we
  construct the rays from geometrical optics.  The result reads
\begin{eqnarray}
\nabla  \times  \nabla\times 
  G(\mathbf{x}_{1},r_{2},\theta_{2},\varphi_{2},t) & = &
 -\frac{\delta^{\prime \prime}(t - \tau + r_{2}/c )} 
 {4\pi c^{2}r_{2}} \, P_{\perp\mathbf{e}_{r_{2}}} \nonumber \\
& \times & 
\begin{cases}
1 &   ( \theta_{2},\varphi_{2} )\in\Omega , \\
0, &  ( \theta_{2},\varphi_{2}) \notin\Omega .
\end{cases}
\label{eq:Green_simple}
\end{eqnarray}
We have introduced the typical time $\tau=(4 f+d)/c$ and we have
  neglected contributions that are small when
$f,d \gg \lambda$. Of course, in Eq.~\eqref{eq:Green_simple} 
we have taken into account that the parabola covers only a finite solid angle 
$\Omega$.

Next, we have to take care of region 3. Here, we use the
free-space propagator, because  the JWKB method would cause a
singularity  at the second focus. The
mentioned propagator for the electric field, which can be derived by
using $G_{\text{free}}$,  is given by 
\begin{eqnarray}
\mathbf{D} ( \mathbf{x}_{2},t)  & = &   \frac{1}{4\pi}
\int_{0}^{\pi} \int _{0}^{2\pi} \sin\theta_{2} 
\left ( \frac{d}{dr_{2}} + \frac{1}{c} \frac{d}{dt} \right ) \, 
 P_{\perp\mathbf{e}_{r_{2}}} \nonumber \\
& \times  &  [ r_{2} \mathbf{D} ( r_{2},\theta_{2},\varphi_{2},
t-\Delta t) ] \, d\theta_{2} d\varphi_{2} \, ,
\end{eqnarray}
where $\Delta t= r_{2}/c$. If we apply this expression
to our problem with the dyadic Green's function we finally obtain
\begin{equation}
\nabla\times\nabla\times G (\mathbf{x}_{1},\mathbf{x}_{2}, t )
= - \frac{1}{3c^{3}\pi} \delta^{\prime \prime \prime} (t-\tau) 
\,  \Gamma_{\text{rel}} \, .
\end{equation}
In the limiting case that the mirrors cover the full solid angle,  we
obtain $\Gamma_{\text{rel}}=\openone$ . 

So far, we have neglected the fact that we could also continue
  the described geometrical rays, which would add further terms to our
  expression for the dyadic Green's function.  But as long we are
only interested in time intervals $t \in[0,2\tau)$ we can neglected
this continuation of the rays. Of course, the relation
\begin{equation}
  \nabla\times\nabla\times G(\mathbf{x}_{2},\mathbf{x}_{1},t)=
  \nabla\times\nabla\times G(\mathbf{x}_{1},\mathbf{x}_{2},t)
\end{equation}
holds true.

We have to calculate also $\nabla\times\nabla\times
G(\mathbf{x}_{1},\mathbf{x}_{1},t)$ and $\nabla\times\nabla\times
G(\mathbf{x}_{2},\mathbf{x}_{2},t)$.  We are retaining only the
dominant contribution, which is associated to the free-space part of
the Green's function. The problem is that this part leads to a divergent
expression that needs to be regularized. For simplicity, we only take
care of the problem $\nabla\times\nabla\times
G_{\text{free}}(\mathbf{x}_{1},r_1,\theta_1,\varphi_1,t)\mathbf{e}_{z}$,
because the symmetry of the problem leads to the general
solution. So,  we get
\begin{equation}
\nabla \times \nabla \times 
G_{\text{free}}(\mathbf{x}_{1}, r_{1}, \theta_{1},\varphi_{1},t)
\mathbf{e}_{z} =\nabla \times \nabla \times
\left [ \frac{\delta (t- r_{1}/c)}{4\pi r_{1}} \right ] \mathbf{e}_{z}
\, .
\end{equation}
In this expression terms of the form $\delta^{(n)} (t-r_{1}/c)/r_{1}^{m}$
appear. By applying a formal Taylor expansion, we get  
\begin{eqnarray}
\frac{\delta^{(n)} (t - r_{1}/c )}{r_{1}^{m}} =
\sum_{k=0}^{\infty} \frac{\delta^{(n+k)} ( t )}{k!} 
r_{1}^{k-m} (-c)^{-k} \nonumber \\
\xrightarrow[r_{1} \rightarrow 0]{} 
\frac{\delta^{(n+m)} (t )}{m!}  (-c )^{-m} +
\sum_{k=0}^{m-1}\frac{\delta^{(n+k)} (t)}{k!}
r_{1}^{k-m} (- c )^{-k} \,. \nonumber \\
\end{eqnarray}
As one can see the sum in the second line contains singular terms,
which lead to the divergent Lamb-shift appearing  after
the dipole approximation. If we neglect those terms, we get 
\begin{equation}
\nabla\times\nabla\times
G_{\text{free}}(\mathbf{x}_{1},\mathbf{x}_{1},t)
 \mathbf{e}_{z} = 
 \frac{\delta^{\prime \prime \prime}(t)}{6\pi c^{3}}\mathbf{e}_{z}\; ,
\end{equation}
which, given the symmetry of the problem, gives the general solution
\begin{equation}
\nabla \times \nabla \times 
G_{\text{free}}(\mathbf{x}_{1},\mathbf{x}_{1},t) = 
\frac{\delta^{\prime \prime \prime}(t)}{6\pi c^{3}} \, .
\end{equation}
From here, we immediately get
\begin{equation}
B_{\alpha;k\ell}^{\alpha;ij} (s)=
\frac{-i  s^{3}}{6\pi c^{3}\epsilon_{0}\hbar} 
\mathbf{d}_{ij}^{\dagger} \mathbf{d}_{k \ell}\, ,
\end{equation}
for $ \alpha\in\left\{ 1,2\right\}$.  Since the time evolution
of the different probability amplitudes is dominated by rapid
oscillations of the form $e^{-i \omega t}$, one can replace each $s$
by $-i \omega$ and we thus get directly the result in
Eq.~\eqref{eq:resfin}.

\section{Tracing out the photonic degrees of freedom}
\label{subTracing_out}

To calculate the fidelity of the entangled state generated by our
scheme, we have to trace out the uncontrolled photonic degrees of
freedom, which in general cause decoherence and destroy entanglement.
Our goal is to determine the reduced density matrix
\begin{eqnarray}
  \hat{\varrho} (t) & = & 
  \Tr_{R} ( | \psi(t) \rangle \langle \psi(t)  |  ) =
  \langle \{ 0 \}  | \psi(t) \rangle \langle \psi(t) | \{0 \} \rangle \nonumber \\
  & + & \sum_{r} \langle 1_{r} |\psi(t) \rangle \langle \psi(t) | 1_{r} \rangle \, .
\end{eqnarray}
The second line of this equation, which denote
$\hat{\varrho}_{\text{ground}}(t)$ is of main interest, because the
ions are affected by spontaneous emission and both of them will be in
the ground state after a short while. To obtain an expression for
$\hat{\rho}_{\text{ground}}(t)$ we have to evaluate an infinite
sum. It is possible to rewrite this sum to a finite sum by using the
functions $A_{\beta,k,l}^{\alpha;i,j} ( s )$.  The result is
the following expression
\begin{widetext}
  \begin{equation}
    \begin{array}{c}
      \bra{i_1^{(1)}}\bra{i_2^{(2)}}\hat{\rho}_{\text{ground}}(t)\ket{j_1^{(1)}}\ket{j_2^{(2)}}
      =\int\limits _{0}^{t}e^{-i
        (t-t')\left(\omega_{i_1}+\omega_{i_2}-\omega_{j_1}-\omega_{j_2}\right)}\\ 
      \left(\sum\limits _{k,l\in S_{\text{e}}}
    \sum\limits_{\alpha\in\{ 1,2 \}} 
\sum\limits_{\beta\in\{ 1,2 \}}\mathcal{L}^{-1}
 \left [ A_{\beta;lj_{\beta}}^{\alpha;ki_{\alpha}} (s+i \omega_{i_1}+i
   \omega_{i_2})
 \widetilde{b}_{ki_{3-\alpha}}^{(\alpha)}(s)\right ] (t^{\prime}) \,
  b_{lj_{3-\beta}}^{(\beta)}[t^{\prime}]^{\ast}\right.\\
      \left.+
 \sum\limits _{k,l\in S_{\text{e}}}
 \sum\limits_{\alpha\in\{ 1,2 \}}
 \sum\limits_{\beta\in\{ 1,2 \}}\mathcal{L}^{-1}
\left[ A_{\alpha;ki_{\alpha}}^{\beta;lj_{\beta}} (s+i \omega_{j_1}+i
  \omega_{j_2})
 \widetilde{b}_{lj_{3-\beta}}^{(\beta)}(s) \right ] [t']^{\ast} \;  
 b_{ki_{3-\alpha}}^{(\alpha)}[t'] \right )dt^{\prime} \, .
    \end{array}
  \end{equation}
\end{widetext}

\bibliography{Literatur}
\end{document}